\documentclass[journal]{IEEEtran}
\usepackage{color}
\usepackage{textcomp}
\usepackage{ifpdf}
\usepackage{cite}
\ifCLASSINFOpdf
   \usepackage[pdftex]{graphicx}
\else
\fi
\usepackage{amsmath}
\usepackage{algorithmic}
\usepackage{array}

\usepackage{stfloats}
\usepackage{url}
\begin{document}
\title{Disordered Anderson Localization Optical Fibers for Image Transport -- A Review}

\author{Arash~Mafi,~\IEEEmembership{Fellow,~OSA,}
        John~Ballato,~\IEEEmembership{Fellow,~OSA,}
        Karl~W.~Koch,~\IEEEmembership{Member,~OSA,}
        and~Axel~Sch\"ulzgen,~\IEEEmembership{Fellow,~OSA}% <-this % stops a space
\thanks{A. Mafi is with the Department of Physics and Astronomy and Center for High Technology Materials, 
University of New Mexico, Albuquerque, NM 87131, USA. e-mail: mafi@unm.edu}% <-this % stops a space
\thanks{J. Ballato is with the Department of Materials Science and Engineering and Center for Optical Materials Science and Engineering Technologies, 
Clemson University, Anderson, SC 29625, USA.}% <-this % stops a space
\thanks{K. W. Koch is with the Applied Optical Physics Directorate, Corning Inc., 
SP-AR-02-4, Sullivan Park, Corning, NY 14831, USA.}
\thanks{A. Sch\"ulzgen is with CREOL, The College of Optics and Photonics, University of Central Florida, Orlando, FL 32816, USA.}
}

\IEEEspecialpapernotice{(Invited Paper)}

\maketitle

\begin{abstract}
Disordered optical fibers show novel waveguiding properties, enabled by the transverse Anderson localization of light, 
and are used for image transport. The strong transverse scattering from the transversely disordered refractive 
index structure results in transversely confined modes that can freely propagate 
in the longitudinal direction. In some sense, an Anderson localization disordered fiber behave like a large-core multimode optical fiber, with the advantage, 
that most modes are highly localized in the transverse plane, so any point in the cross section of the fiber can be 
used for localized beam transport. This property has been used for high-quality transportation of intensity patterns and images  
in these optical fibers. This review covers the basics and the history of the transverse Anderson localization in disordered optical fibers
and captures the recent progress in imaging applications using these optical fibers.
\end{abstract}

%\begin{IEEEkeywords}
%Optical Fibers, Optical fiber radiation effects, CW lasers
%\end{IEEEkeywords}

% \fi
\IEEEpeerreviewmaketitle

\section{Introduction}
%%%%%%%%%%%%%%%%%%%%%%%%%%%%%%%%%%%%%%%%%%%%%%%%%%%%%%%%%%%%%%%%%%%%%%%
\IEEEPARstart{A}{nderson} localization (AL) is the absence of diffusive wave transport in highly disordered 
scattering media~\cite{Anderson1,Anderson1980,Abrahams-50-book,Lagendijk-Physics-Today-2009,Abrahams-Scaling-Theory,Stone,sheng2006introduction}. 
It is broadly applicable to the quantum mechanical wave
function of an electron described by the Schr\"odinger equation~\cite{Anderson1,Thouless-1974,Wegner1976,Soukoulis-1999}, 
matter waves and Bose-Einstein condensates~\cite{matter-waves-2008,roati2008anderson,kondov2011three}, 
quantum fields such as photons in various quantum optical systems~\cite{quantum-fields-2010,Lahini-Quantum-Correlation-2010,Lahini-HBT-2011,Abouraddy-entangled-2012}, 
as well as classical wave phenomena including acoustics~\cite{ultrasound-1990,acoustic-PRL-1990}, 
elastics~\cite{elastics-Nat-Phys-2009}, electromagnetics~\cite{John-EM-abs-mobility-edge-1984,dalichaouch1991microwave,Chabanov-microwave-2000,El-Dardiry-microwave-2012}, 
and optics~\cite{Anderson2,John-photon-localization-1987,John-Physics-Today-1991,SegevNaturePhotonicsReview,Mafi-AOP-2015}. 

Among all classical wave systems, optics is uniquely positioned for studies of AL phenomena because of the
 diverse set of possibilities to construct the disordered background potential and the availability of robust tools to 
experiment and probe the localization phenomena~\cite{storzer2006observation,wiersma1997localization,yannopapas2003anderson,aegerter2007observation},
including its behavior in the presence of 
nonlinearity~\cite{Lahini-1D-AL-2008,fishman2012nonlinear,mafi-NL-ArXiv-2017,Mafi-Marco-PRL-Migrating-NL-2014,Mafi-Marco-APL-self-focusing-2014}.
There have been many attempts over the years to observe AL of light in a three-dimensional (3D) disordered optical medium, 
including some recent demonstrations~\cite{sperling2013direct,vatnik2017anderson,choi2018anderson}. However, because the large refractive index contrasts required
for 3D Anderson localization are generally accompanied with considerable losses in optics, and because it is not easy to differentiate between the exponential
decay of the optical field associated with loss and the exponential decay of the field due to the AL, 3D AL of light remains a 
subject of active on-going research.

In this Review, our focus is on the transverse Anderson localization (TAL) of light in a waveguide-like structure. In TAL structures, 
the dielectric constant is uniform along the direction of the propagation of light, similar to a conventional optical waveguide,
and the disorder in the dielectric constant resides in the (one or two) transverse dimension(s). An optical 
field that is launched in the longitudinal direction tends to remain localized in the disordered transverse dimension(s) but
propagates freely as if in an optical waveguide in the longitudinal direction. TAL appears to be ubiquitous in any transversely disordered
waveguide, as long as the disorder is sufficiently strong such that the transverse physical dimensions of the waveguide are larger than the
transverse localization length and the waveguide remains uniform in the longitudinal direction. 
In the following, after a brief historical survey of the origins of the TAL, we will present the recent progress
especially as related to TAL in disordered optical fibers with particular emphasis on applications to image transport and incoherent illumination.
%%%%%%%%%%%%%%%%%%%%%%%%%%%%%%%%%%%%%%%%%%%%%%%%%%%%%%%%%%%%%%%%%%%%%%%%%%%%%%%%
%%%%%%%%%%%%%%%%%%%%%%%%%%%%%%%%%%%%%%%%%%%%%%%%%%%%%%%%%%%%%%%%%%%%%%%%%%%%%%%%
\section{Brief Survey on the Origins of TAL}
%%%%%%%%%%%%%%%%%%%%%%%%%%%%%%%%%%%%%%%%%%%%%%%%%%%%%%%%%%%%%%%%%%%%%%%%%%%%%%%%
%%%%%%%%%%%%%%%%%%%%%%%%%%%%%%%%%%%%%%%%%%%%%%%%%%%%%%%%%%%%%%%%%%%%%%%%%%%%%%%%
Transverse Anderson localization in a quasi-2D optical system was first proposed in a pair of visionary theoretical papers by 
Abdullaev \textit {et al}.~\cite{transverse-Abdullaev} in 1980 and De~Raedt \textit {et al}.~\cite{transverse-DeRaedt} in 1989.
%%%%%%%%%%%%%%%%%%%%%%%%%%%%%%%%%%%%%%%%%%%%%%%%%%%%
\begin{figure}[htp]
  \centering
  \includegraphics[width=0.9\columnwidth]{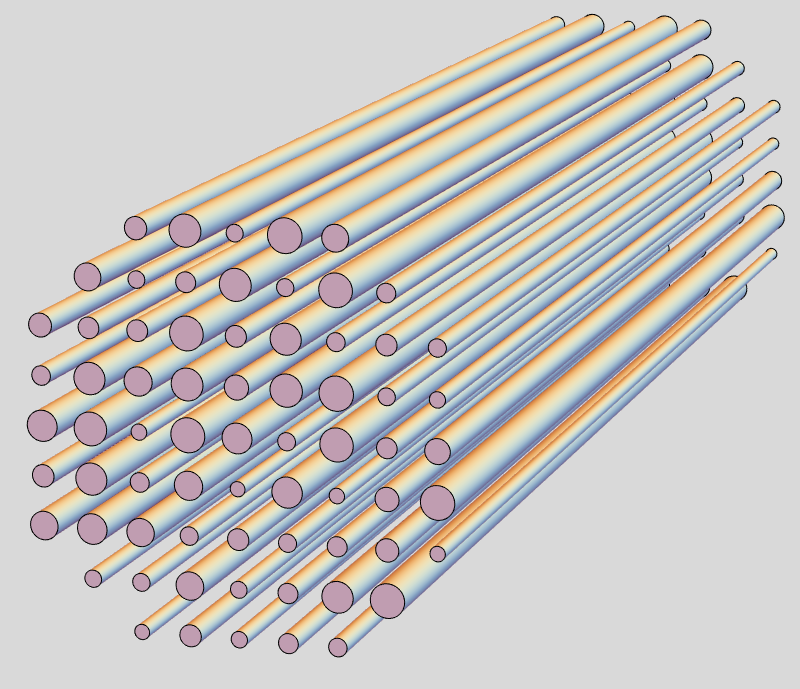}
\caption {A conceptual sketch of the 2D randomized array of coupled optical fibers is shown, as proposed by 
Abdullaev \textit {et al}.~\cite{transverse-Abdullaev}, to observe the TAL of light.}
\label{fig:abd-array}
\end{figure}
%%%%%%%%%%%%%%%%%%%%%%%%%%%%%%%%%%%%%%%%%%%%%%%%%%%%
The structure proposed by Abdullaev \textit {et al}.~\cite{transverse-Abdullaev} is sketched in Fig.~\ref{fig:abd-array}, 
consisting of a two-dimensional (2D) array of coupled optical fibers with slightly different and randomly distributed physical parameters, 
e.g., different radii. Therefore, the propagation constants of the guided modes supported by the optical fibers are randomly distributed.
Because the individual fibers are evanescently coupled, light is expected to tunnel from one fiber to another. However, the
efficiency of the optical tunneling between neighboring fibers is reduced because the propagation constants of the modes 
are generally different due to the randomness~\cite{Saleh-Teich}. Therefore, if the light is coupled initially in one optical fiber, 
it does not spread out as efficiently to other fibers and the amplitude of the field, on average, decays exponentially in the 
transverse dimensions.
This localization is readily observable if the nominal transverse decay length (localization radius) is 
smaller than the transverse dimensions of the system. Of course, the localization radius can generally be made smaller by increasing
the randomness.  

%%%%%%%%%%%%%%%%%%%%%%%%%%%%%%%%%%%%%%%%%%%%%%%%%%%%
\begin{figure}[htp]
  \centering
  \includegraphics[width=0.9\columnwidth]{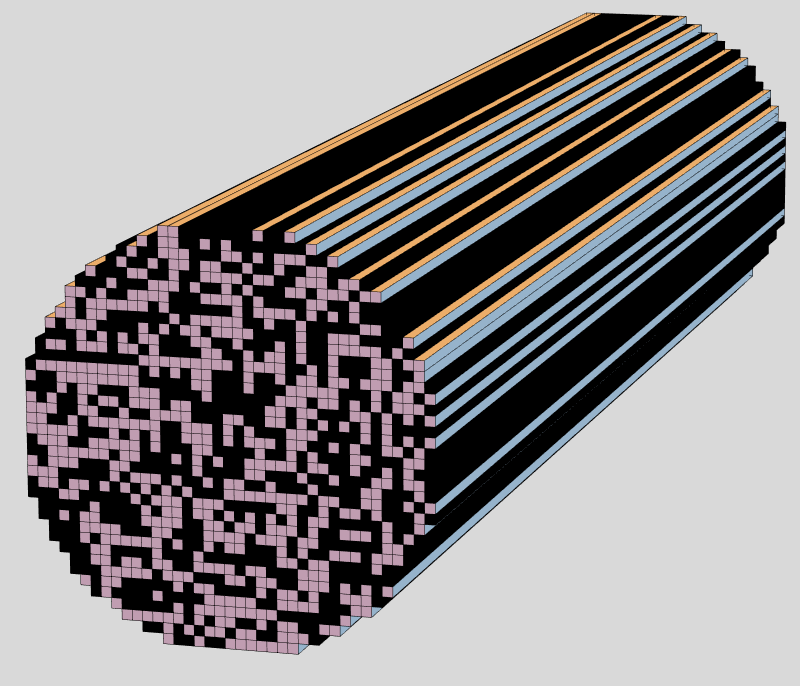}
\caption {A conceptual sketch of the transversely random and longitudinally invariant dielectric
medium for the observation of the TAL of light is shown, as proposed by De~Raedt \textit {et al}.~\cite{transverse-DeRaedt}.
The refractive index is invariant in the longitudinal direction. In the transverse plane, the refractive index is 
pixelated into tiny squares, and the refractive index of each pixel is randomly selected to be $n_1$ or $n_2$ with equal probabilities.
}
\label{fig:rae-array}
\end{figure}
%%%%%%%%%%%%%%%%%%%%%%%%%%%%%%%%%%%%%%%%%%%%%%%%%%%%
The structure proposed by De~Raedt \textit {et al}.~\cite{transverse-DeRaedt} is sketched in Fig.~\ref{fig:rae-array},
consisting of an optical fiber-like structure, whose refractive index profile is invariant in the longitudinal direction. In the
transverse plane, the refractive index is pixelated into tiny squares, where the edge length of each square is on the order of 
the wavelength of the light. The refractive index of each pixel is randomly selected to be $n_1$ or $n_2$ with equal probabilities.
De~Raedt \textit {et al}. showed that an optical field that is launched in the longitudinal direction tends to remain localized 
in the transverse plane due to the transverse scattering and the amplitude of the field, on average, decays exponentially in the 
transverse dimensions, as it propagates freely in the longitudinal direction. The 
localization radius can be generally reduced by increasing the refractive index contrast $\Delta n=|n_2-n_1|$. 

One of the earliest experimental attempts to observe TAL was carried out by Pertsch \textit {et al}.~\cite{Pertsch},
where they investigated light propagation in a disordered 2D array of mutually coupled optical fibers, similar to the structure proposed 
by Abdullaev \textit {et al}.~\cite{transverse-Abdullaev}. They made interesting observations in the nonlinear regime, where they showed 
that for high excitation power, diffusive spreading is arrested by the focusing nonlinearity. However, the disorder was not sufficiently 
large in their structure to result in the observation of TAL. In other words, the localization radius in their structure appears to have been larger than
the transverse dimensions of their 2D array. 

The first successful attempt in observing TAL was reported by Schwartz \textit {et al}.~\cite{Schwartz2007} and was performed in a   
photorefractive crystal. An intense laser beam was used to write the transversely 
disordered and longitudinally invariant refractive index profiles in a photorefractive crystal, 
and another laser probe beam was used to investigate the transverse localization behavior. Their experiment allowed 
them to vary the disorder level by controlling the laser illumination of the photorefractive crystal in a controlled
fashion to observe the onset of the transverse localization and the changes in the localization 
radius as a function of the disorder level. Because the variations in the refractive index of the random sites 
were on the order of $10^{-4}$, the localization radius was observed to be considerably larger than the wavelength
of the light. 

Over the next few years after the pioneering demonstration by Schwartz \textit {et al}.~\cite{Schwartz2007}, several
theoretical and experimental efforts in one-dimensional (1D) disordered lattices were performed that demonstrated
and further explored various aspects of the TAL phenomena, including the impact of the Kerr nonlinearity 
and boundary effects~\cite{Lahini-1D-AL-2008,Szameit-boundary-2010,Martin-1D-AL-2011,Kartashov-NL-2012}. These efforts eventually
led to the development of TAL optical fibers that will be discussed in the rest of the Review.
%%%%%%%%%%%%%%%%%%%%%%%%%%%%%%%%%%%%%%%%%%%%%%%%%%%%%%%%%%%%%%%%%%%%%%%%%%%%%%%%
%%%%%%%%%%%%%%%%%%%%%%%%%%%%%%%%%%%%%%%%%%%%%%%%%%%%%%%%%%%%%%%%%%%%%%%%%%%%%%%%
\section{TAL in Disordered Optical Fibers}
\begin{figure}[t]
  \centering
  \includegraphics[width=0.9\columnwidth]{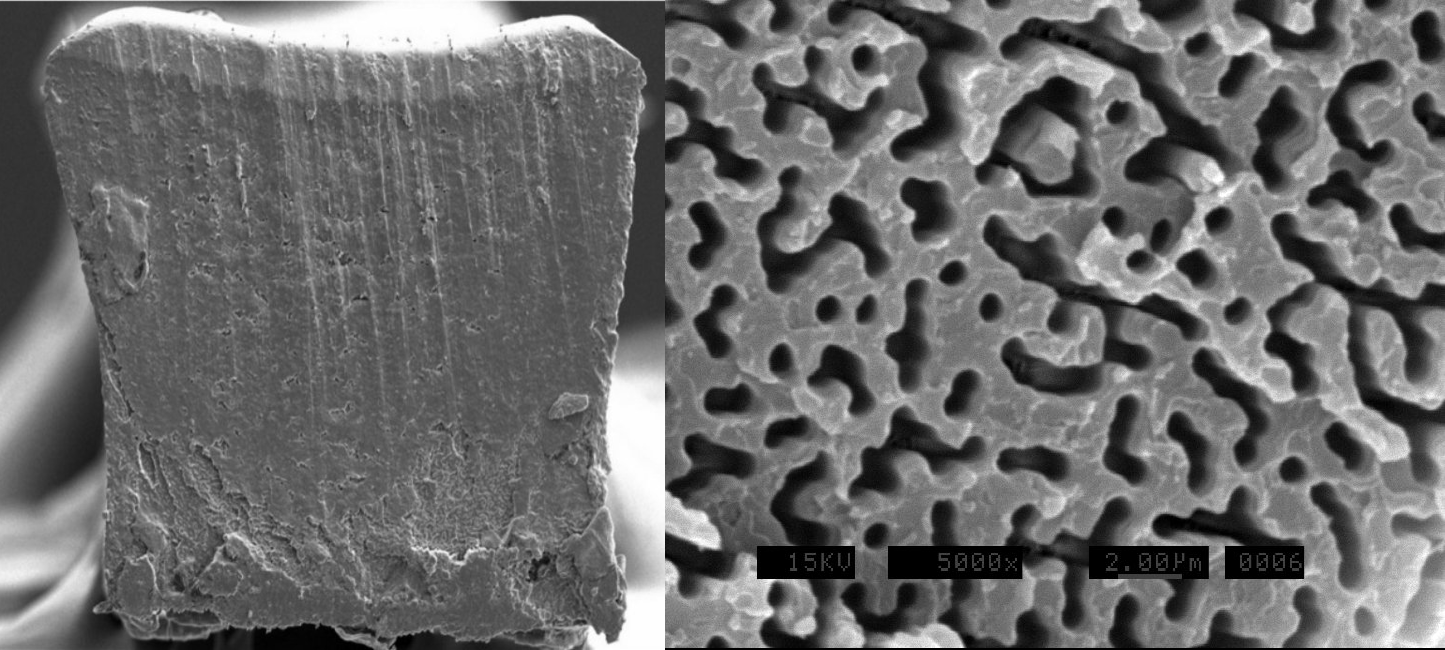}
\caption {Cross section of the polymer disordered fiber from Ref.~\cite{Mafi-Salman-OL-2012}
is shown with a nearly square profile and an approximate side-width of 250\textmu m in the left panel.
A zoomed-in SEM image of a 24\textmu m-wide region on the tip of the fiber, exposed to a solvent 
to differentiate between PMMA and PS polymer components, is shown in the right panel.
Feature sizes are around 0.9\textmu m, and darker regions are PMMA. Reprinted/Reproduced
with permission from Optics Letters, 2012~\cite{Mafi-Salman-OL-2012}, and the Optical Society of America.}
\label{fig:karbasi-polymer}
\end{figure}

%%%%%%%%%%%%%%%%%%%%%%%%%%%%%%%%%%%%%%%%%%%%%%%%%%%%%%%%%%%%%%%%%%%%%%%%%%%%%%%%
%%%%%%%%%%%%%%%%%%%%%%%%%%%%%%%%%%%%%%%%%%%%%%%%%%%%%%%%%%%%%%%%%%%%%%%%%%%%%%%%
The first demonstration of TAL in an optical fiber was reported in 2012 by Karbasi \textit {et al}.~\cite{Mafi-Salman-OL-2012}.
The structure used by Karbasi \textit {et al}. is shown in Fig.~\ref{fig:karbasi-polymer}, which is similar to the design proposed by
De~Raedt \textit {et al}.~\cite{transverse-DeRaedt}. The optical fiber was fabricated by the stack-and-draw method
from a low-index component, polymethyl methacrylate (PMMA) with a refractive index of 1.49, and a high-index component,
polystyrene (PS) with a refractive index of 1.59. 40,000~pieces of PMMA and 40,000 pieces of PS fibers were 
randomly mixed~\cite{Mafi-Salman-JOVE-2013}, fused together, and redrawn to a fiber with a nearly square profile and 
approximate side-width of 250\textmu m, as shown in the left panel in Fig.~\ref{fig:karbasi-polymer}.
The right panel shows the zoomed-in scanning electron microscope~(SEM) image of an approximately 24\textmu m-wide region on 
the tip of the fiber after exposing the tip to an ethanol solvent to dissolve the PMMA. The typical random feature size 
in the structure shown in Fig.~\ref{fig:karbasi-polymer} is around 0.9\textmu m.

Karbasi \textit {et al}. demonstrated that when the light was launched into the disordered fiber
from a small-core single-mode optical fiber, the beam went through a brief initial expansion
(transverse diffusion) but the expansion was arrested upon propagating for $\sim$2cm, after which
the TAL eventually took over. The mean effective beam radius for the 100 measured near-field beam 
intensity profiles was calculated to be $\xi_{\rm avg}=31$\textmu m, with a standard deviation
$\sigma_{\xi}=14$\textmu m. TAL was observed in samples as long as 60cm, but the large variations 
in the thickness of the optical fiber in the draw process hindered the observation of TAL 
in longer samples. Furthermore, it was observed that when the input beam was 
scanned across the input facet, the output beam followed the transverse position of the 
incoming beam~\cite{Mafi-Salman-OPEX-2012}. 
%%%%%%%%%%%%%%%%%%%%%%%%%%%%%%%%%%%%%%%%%%%%%%%%%%%%
%%%%%%%%%%%%%%%%%%%%%%%%%%%%%%%%%%%%%%%%%%%%%%%%%%%%

Subsequently, further detailed analyses of the disordered fibers in Ref.~\cite{Mafi-Salman-OL-2012} were
conducted in Ref.~\cite{Mafi-Salman-OPEX-2012} to explore the effect of the refractive index contrast,
fill-fraction, and random site size on the localization radius. It was shown that at least for $\Delta n\le 0.5$,
the larger index contrast results in a stronger AL and smaller localization radius. However, the jury is 
still out for larger values of the index contrast, especially if $\Delta n$ becomes so large that the vectorial 
nature of the optical field must be taken into account~\cite{Skipetrov}. 
The optimal value of the fill fraction, defined as the fraction of the low-index polymer to the total, 
was shown to be 50\%, resulting in the strongest transverse scattering. It is notable that the optimal 
50\% value is below the percolation threshold (59.27\%) of a square lattice; therefore, the host
material with the higher refractive index remains generally connected in the long range, making the AL
non-trivial, i.e., it is not merely due to the disconnected clusters of the higher index material. 

The initial studies on the impact of the random site size showed that the edge length of each square pixel must ideally
be equal to half the wavelength of the light. However, this observation was contradicted in later 
studies, as will be discussed in more detail in subsection~\ref{sec:optimal}.
Another important observation made in Ref.~\cite{Mafi-Salman-OPEX-2012} was that the statistical distribution 
of the mode field diameters follows a nearly Poisson-like distribution, i.e., a stronger TAL that leads to a 
smaller average mode field diameter also reduces the mode-to-mode diameter variations; therefore, 
a stronger TAL leads to a stronger uniformity in the supported modes across the disordered fiber. These observations 
resulted in the understanding that a stronger AL, especially using higher index contrast components, is warranted, which 
eventually led to the development of a glass-air disordered fiber structure.  

\begin{figure}[t]
  \centering
  \includegraphics[width=0.9\columnwidth]{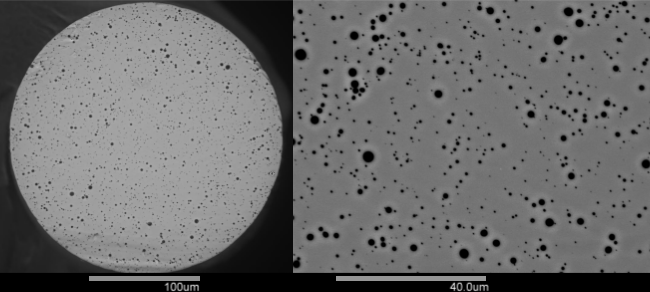}
\caption {SEM image of the glass optical fiber with random airholes reported in
Ref.~\cite{Mafi-Salman-OMEX-2012} is shown in the left panel. A zoomed-in SEM image of the same fiber
is shown in the right panel. Reprinted/Reproduced
with permission from Optical Material Express, 2012~\cite{Mafi-Salman-OMEX-2012}, and the Optical Society of
America.}
\label{fig:SEM-Magnified-2}
\end{figure}

The first observation of the TAL in a silica fiber was reported by Karbasi \textit {et al}.~\cite{Mafi-Salman-OMEX-2012} 
in~2012. The glass-air disordered fiber used for this work was drawn at Clemson University, where the preform was made from 
``satin quartz'' (Heraeus Quartz), which is a porous artisan glass. The airholes in the porous glass were drawn to
air channels; therefore, the structure resembled the design proposed by De~Raedt \textit {et al}.~\cite{transverse-DeRaedt},
where $n_1=1.0$ and $n_2=1.46$. The cross-sectional SEM image of this fiber is shown in Fig.~\ref{fig:SEM-Magnified-2}
in the left panel, and a zoomed-in SEM image is shown in the right panel. The light-gray background matrix is glass, and the 
random black dots represent the airholes. The total diameter of the disordered glass-air fiber was measured to be 250\textmu m. 
The diameters of the airholes varied between 0.2\textmu m and 5.5\textmu m. We note that the fill-fraction of the airholes in this fiber
ranged from nearly 2\% in the center of the fiber to approximately 7\% near the edges; therefore, TAL was only 
observed near the periphery of the fiber. This caused a bit of debate, considering the perceived delocalizing impact of the boundaries
in disordered TAL systems~\cite{Szameit-boundary-2010,Jovic-boundary-2011,Molina-boundary-2011,Naether-boundary-2012}, which was subsequently addressed
in Ref.~\cite{Mafi-Behnam-Boundary-OC-2016}, as will be discussed in more detail in subsection~\ref{sec:boundary}.     
%%%%%%%%%%%%%%%%%%%%%%%%%%%%%%%%%%%%%%%%%%%%%%%%%%%%
%%%%%%%%%%%%%%%%%%%%%%%%%%%%%%%%%%%%%%%%%%%%%%%%%%%%

In 2014, there was another successful attempt in observing TAL in an air-glass optical fiber by Chen and Li~\cite{chen2014observing} at
Corning Incorporated.
They fabricated random air-line fibers with approximately 150, 250 and 350\,\textmu m diameters and observed TAL with 
significantly lower air fill-fraction than those reported in Ref.~\cite{Mafi-Salman-OMEX-2012}. This can be attributed to the
far-subwavelength size of the transverse scattering centers and the higher scattering center density (air-line density) than
the fiber studied in Ref.~\cite{Mafi-Salman-OMEX-2012}. There have since been other successful attempts is observing TAL in glass-based
fibers, such as the air-silica random fiber structure by Zhao \textit {et al}.~\cite{ZHAO:17,zhao2018image} at CREOL, University of Central Florida,
and the fabrication of an all-solid tellurite optical glass by Tuan \textit {et al}.~\cite{tong2018characterization} from the
Toyota Technological Institute in Japan. These recent reports will be discussed in more detail in section~\ref{sec:imaging}. 
%%%%%%%%%%%%%%%%%%%%%%%%%%%%%%%%%%%%%%%
%%%%%%%%%%%%%%%%%%%%%%%%%%%%%%%%%%%%%%%
\subsection{Optimal pixel size and wavelength dependence of TAL}
\label{sec:optimal}
%%%%%%%%%%%%%%%%%%%%%%%%%%%%%%%%%%%%%%%
%%%%%%%%%%%%%%%%%%%%%%%%%%%%%%%%%%%%%%%
The issue of the optimal pixel size and the wavelength dependence of TAL were initially explored in Ref.~\cite{Mafi-Salman-OPEX-2012}.
For the structure proposed by De~Raedt \textit {et al}.~\cite{transverse-DeRaedt}, because Maxwell's equations 
are scale invariant, {\em increasing the pixel size while keeping the wavelength fixed} can be trivially mapped to
{\em decreasing the wavelength while keeping the pixel size fixed}. It was initially claimed that a
shorter wavelength (or equivalently a larger pixel size at a given wavelength) decreased the localization 
radius (in units of the pixel size)~\cite{Mafi-Salman-OPEX-2012}.
However, further analysis and subsequent work in 1D~\cite{Mafi-Salman-Modal-JOSAB-2013} hinted that the optimum value of the pixel
size is around half the free-space wavelength, at least for the refractive index on the order of 1.5. However, more recent 
experimental evidence and simulations in Ref.~\cite{Mafi-Schirmacher-PRL-2018} have cast doubt on this observation. 
Schirmacher \textit {et al}.~\cite{Mafi-Schirmacher-PRL-2018} argued that the average localization radius shows 
no dependence on the wavelength (over a reasonable range). They attributed the observation of the wavelength dependence for the simulations 
presented in Ref.~\cite{Mafi-Salman-OPEX-2012} to the omission of a term proportional to the gradient of the dielectric permittivity,
which is common in the finite difference simulations of optical fibers but not acceptable for TAL fiber.
They also noted that the large error bars in the experiments performed in Ref.~\cite{Mafi-Salman-OPEX-2012} may have been behind the 
disagreements in the experimental observations; however, simulations in Ref.~\cite{Mafi-Salman-Modal-JOSAB-2013}
correctly took the permittivity gradient term into account, and still showed some wavelength dependence. 
Therefore, the issue is not entirely settled, and part of the disagreement may reside in different ways the averaging is performed. 
As of now, the jury is still out, and this issue needs to be explored in further detail.    
%%%%%%%%%%%%%%%%%%%%%%%%%%%%%%%%%%%%%%%
%%%%%%%%%%%%%%%%%%%%%%%%%%%%%%%%%%%%%%%
\subsection{TAL near the disordered fiber boundaries}
\label{sec:boundary}
%%%%%%%%%%%%%%%%%%%%%%%%%%%%%%%%%%%%%%%
%%%%%%%%%%%%%%%%%%%%%%%%%%%%%%%%%%%%%%%
TAL of light near the boundaries was discussed theoretically in Refs.~\cite{Jovic-boundary-2011,Molina-boundary-2011} and experimentally in
Refs.~\cite{Szameit-boundary-2010,Naether-boundary-2012}. They originally reported a delocalizing effect near the boundaries of 1D and 2D random lattice 
waveguides. These reports appeared to be in contrast with the experimental observation reported by Karbasi \textit {et al}.~\cite{Mafi-Salman-OMEX-2012}, 
who reported that a strong localization happened near the outer boundary of the glass-air disordered fiber and no trace of localization was observed in the
central regions. The disagreements were explained in Ref.~~\cite{Mafi-Salman-OMEX-2012} by 
the non-uniform distribution of disorder in the fiber, where the disorder
was measured to be much stronger near the outer boundary of the fiber, which resulted in a stronger localization in that region. However, 
Abaie \textit {et al}. later performed a detailed analysis in Ref.~\cite{Mafi-Behnam-Boundary-OC-2016} and showed that the perceived
suppressed localization near the boundaries is due to a lower mode density near the boundaries compared with
the bulk, while the average decay rate of the tail of localized modes is the same near the boundaries as in bulk. Therefore,
on average, it is less probable to excite a localized mode near the boundaries; however, once it is excited, its localization is with the same 
exponential decay rate as any other localized mode.
%%%%%%%%%%%%%%%%%%%%%%%%%%%%%%%%%%%%%%%%%%%%%%%%%%%%%%%%%%%%%%%%%%%%%%%%%%%%%%%%
%%%%%%%%%%%%%%%%%%%%%%%%%%%%%%%%%%%%%%%%%%%%%%%%%%%%%%%%%%%%%%%%%%%%%%%%%%%%%%%%
\section{Image Transport and Illumination using Disordered Optical Fibers}
\label{sec:imaging}
\begin{figure}[t]
  \centering
  \includegraphics[width=0.5\columnwidth]{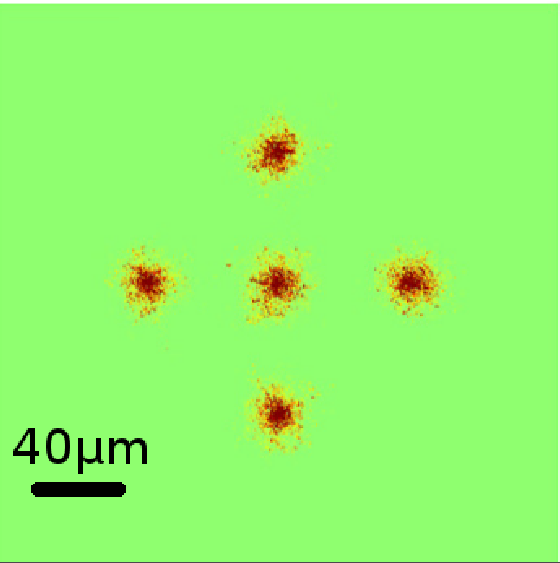}
\caption {The image shows the transportation of five multiplexed beam (computational) in the disordered polymer 
of Ref.~\cite{Mafi-Salman-OL-2012} for a propagation distance of 5cm at 405nm wavelength, where no interference between the beams 
is observed in the output. The green background area represents the fiber area with a side-width of 250\textmu m.
Reprinted/Reproduced with permission from Optics Express, 2012~~\cite{Mafi-Salman-Multiple-Beam-2013}, and the Optical Society of
America.}
\label{fig:five-beams}
\end{figure}

%%%%%%%%%%%%%%%%%%%%%%%%%%%%%%%%%%%%%%%%%%%%%%%%%%%%%%%%%%%%%%%%%%%%%%%%%%%%%%%%
%%%%%%%%%%%%%%%%%%%%%%%%%%%%%%%%%%%%%%%%%%%%%%%%%%%%%%%%%%%%%%%%%%%%%%%%%%%%%%%%
Once the localized beam propagation was verified in highly disordered multi-component polymer and air-glass fibers, the next natural 
step was to explore the possibility of beam multiplexing in these fibers. This was reported in Ref.~\cite{Mafi-Salman-Multiple-Beam-2013},
where Karbasi \textit {et al}. investigated the simultaneous propagation of multiple beams in a disordered TAL fiber. Moreover,
it was shown that the multiple-beam propagation was quite robust to macro-bending and even a tight bending radius in the range of 
2-4mm did not result in any notable beam drifts and the multi-beam structure remained intact. In Fig.~\ref{fig:five-beams},
we show an example of the multibeam propagation in the disordered polymer of Ref.~\cite{Mafi-Salman-OL-2012} for a propagation
distance of 5cm at 405nm wavelength.
%%%%%%%%%%%%%%%%%%%%%%%%%%%%%%%%%%%%%%%%%%%%%%%%%%%%
%%%%%%%%%%%%%%%%%%%%%%%%%%%%%%%%%%%%%%%%%%%%%%%%%%%%
Motivated by the successful demonstration of beam multiplexing, Karbasi \textit {et al}.~\cite{Mafi-Salman-Image-2013} 
compared the quality of image transport in a 1D waveguide with a periodic structure to the image transport in a
disordered waveguide. The periodic waveguide was meant as a 1D prototype of a coherent fiber optic bundle that is commonly 
used for imaging applications. It was shown that increased disorder improved the quality of the image transport.

\begin{figure}[t]
  \centering
  \includegraphics[width=0.9\columnwidth]{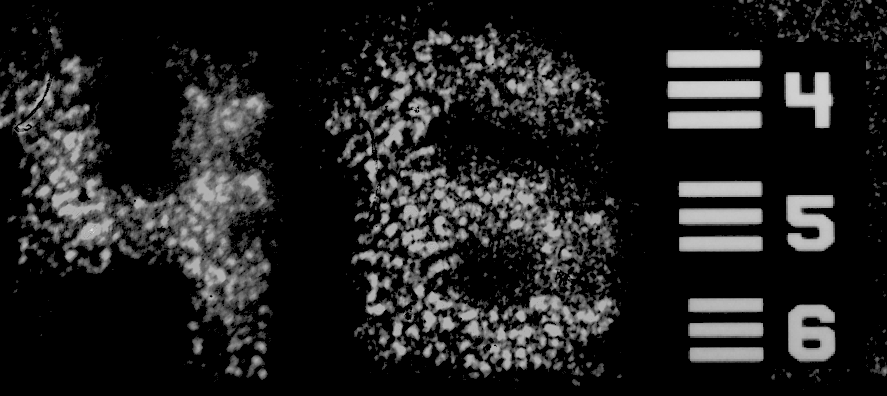}
\caption {Transported images of numbers ``4'' and ``6'' through a 5cm piece of a disordered fiber are shown 
and a section of the 1951 U.S. Air Force resolution test chart (1951-AFTT) used in the image transport 
experiment is shown. The images are 120\textmu m long. Details can be obtained in Reference~\cite{Mafi-Salman-Image-2013}.}
\label{fig:image-transport-2}
\end{figure}

In a subsequent study reported in Ref.~\cite{Mafi-Salman-Nature-2014}, Karbasi \textit {et al}. explored image propagation in
the TAL polymer fiber of Ref.~\cite{Mafi-Salman-OL-2012}. They showed that the image transport quality was comparable with 
or better than some of the best commercial multicore imaging fibers, with less pixelation and higher contrast.
Figure~\ref{fig:image-transport-2} shows an example of a transported image in the form of numbers from a section of the 
1951 U.S. Air Force resolution test chart through the disordered fiber. The test-target, a section of which is shown in the right panel of 
Fig.~\ref{fig:image-transport-2}, is in the form of a stencil in which 
numbers and lines are carved--it was butt-coupled to the hand-polished input facet of the fiber and was illuminated by white light.
The near-field output was projected onto a CCD camera with a 40$\times$ microscope objective.

The minimum resolution of the images is determined by the width of the point-spread function of the disordered optical
fiber imaging system (localization radius), which was calculated to be smaller than 10\textmu m at 405nm wavelength~\cite{Mafi-Salman-OPEX-2012}.
The perceived image quality of the transported images was quantified by the mean structural similarity index (MSSIM), using 
which it was verified that a disordered TAL can transport images of higher quality than conventional coherent fiber bundles.
%%%%%%%%%%%%%%%%%%%%%%%%%%%%%%%%%%%%%%%%%%%%%%%%%%%%
%%%%%%%%%%%%%%%%%%%%%%%%%%%%%%%%%%%%%%%%%%%%%%%%%%%%
Another notable work using the disordered polymer fibers of Ref.~\cite{Mafi-Salman-OL-2012} was the demonstration by
Leonetti \textit {et al}.~\cite{Mafi-Marco-information-2016} of propagating reconfigurable localized optical patterns 
in the fiber to encode up to 6 bits of information in disorder-induced high transmission channels, even using a small 
number of photon counts. This effort highlighted the potential application of these fibers in quantum information processing
in spatially multiplexed configurations.

The first successful image transport in an air-silica random fiber structure was reported in 2018 by
Zhao \textit {et al}.~\cite{ZHAO:17,zhao2018image}, where the disordered fiber featured a 28.5\% air-filling 
fraction in the structured region, and low attenuation below 1dB per meter at visible wavelengths. 
High-quality optical image transfer through 90 cm-long fibers was reported in these disordered fibers. 
In a more recent attempt, Zhao \textit {et al}.~\cite{zhao2018deep} applied deep learning techniques
to improve the quality of the image transport in these fibers. The system they have developed provides the unique 
property that the training performed within a straight fiber setup can be utilized for high fidelity reconstruction 
of images that are transported through either straight or bent fiber, hence making the retraining for different bending 
situations unnecessary. This report is a considerable advancement compared with previous demonstrations of image transport 
in multimode fiber, such as the report by Choi \textit {et al}.~\cite{Choi-multimode}, where the computed
transmission matrix had to be recalculated for any bending or twisting the fiber, making the method slow
and computationally very challenging.

Most recently, Tuan \textit {et al}.~\cite{tong2018characterization} reported the fabrication of the first 
all-solid tellurite optical glass rod with a transversely disordered refractive index profile and a
refractive index contrast of $\Delta n=$0.095 to study TAL of light and near-infrared (NIR) optical image transport.
Experiments performed at the NIR optical wavelength of 1.55\textmu m  confirmed TAL in this structure,
and the images transported over a 10cm length of the disordered fiber showed high contrast and high brightness.
Last but not least, we would like to highlight the work led by Thomas P Seward III of Corning Incorporated
in the 1970s on phase-separated glasses that resulted in random elongated needle-like structures
after drawing~\cite{seward1974elongation,seward1977some}. The fiber-like glass rods were successfully used 
for image transport and most likely operated based on the TAL principles discussed here.
%%%%%%%%%%%%%%%%%%%%%%%%%%%%%%%%%%%%%%%
%%%%%%%%%%%%%%%%%%%%%%%%%%%%%%%%%%%%%%%
\subsection{Mode-area probability density function and scaling}
%%%%%%%%%%%%%%%%%%%%%%%%%%%%%%%%%%%%%%%
%%%%%%%%%%%%%%%%%%%%%%%%%%%%%%%%%%%%%%%
Scaling properties of TAL structures can provide a wealth of information on their physical 
properties~\cite{Anderson1980,Abrahams-Scaling-Theory,Wegner1976,Stone,Pichard1981-1,Pichard1981-2,Pichard1986-1,Pichard1986-2,Aegerter:07}.
We briefly discussed the issue related to the optimal pixel size or the imaging wavelength in TAL fibers in section~\ref{sec:optimal}. 
We argued that some of the discrepancies might reside in different ways the averaging is performed. There is one more critical issue that
 must be considered before making broad-reaching conclusions. The traditional study of TAL is based on launching a beam and analyzing 
its propagation along the waveguide. It is hard to ensure that the results are not biased by the choice of the initial
launch condition. To address this issue, Abaie \textit {et al}.~\cite{Mafi-Behnam-Scaling-PRB-2016,Mafi-Abaie-OL-2018}
performed detailed studies on quasi-1D and quasi-2D (fiber-like) TAL structures using the modal method and calculated the   
mode-area (MA) probability density function (PDF) for these structures. The MA-PDF encompasses all the relevant statistical information
on TAL; it relies solely on the physics of the disordered system and the physical nature of the propagating wave 
and is independent of the beam properties of the external excitation. For their analysis in Ref.~\cite{Mafi-Abaie-OL-2018},
Abaie \textit {et al}. used a quasi-2D structure that was based on the 
random fiber design proposed by De~Raedt \textit {et al}.~\cite{transverse-DeRaedt}. 

Although Refs.~\cite{Mafi-Behnam-Scaling-PRB-2016,Mafi-Abaie-OL-2018}
provide a wealth of information on the inner workings of the TAL behavior, especially when it comes to differentiating between the localized and extended
modes and the best strategies to optimization of the waveguide, they have yet to be adequately leveraged to address the discrepancies discussed
in section~\ref{sec:optimal}. A key observation reported in Refs.~\cite{Mafi-Behnam-Scaling-PRB-2016,Mafi-Abaie-OL-2018}
was that the MA-PDF can be reliably computed from structures with substantially smaller transverse dimensions than the size of the practical waveguides
used in experiments. In fact, it is shown that the shape of the MA-PDF rapidly converges to a terminal form as a function
of the transverse dimensions of the waveguide. This notable scaling behavior observed in MA-PDF is of immense practical
importance in the design and optimization of such TAL-based waveguides, because one can obtain all the useful TAL
information from disordered waveguides with smaller dimensions, hence substantially reducing the computational cost.
%%%%%%%%%%%%%%%%%%%%%%%%%%%%%%%%%%%%%%%
%%%%%%%%%%%%%%%%%%%%%%%%%%%%%%%%%%%%%%%
\subsection{Spatial coherence and illumination}
%%%%%%%%%%%%%%%%%%%%%%%%%%%%%%%%%%%%%%%
%%%%%%%%%%%%%%%%%%%%%%%%%%%%%%%%%%%%%%%
Although image transport has so far been the main focus of the research efforts on TAL fibers, control of spatial coherence 
and illumination are also potentially viable areas that are due for further explorations. 
In particular, Refs.~\cite{Mafi-Marco-singlemode-2017,Mafi-Behnam-Optica-2018} recently pointed out that the presence of a strong 
disorder in TAL fibers can be exploited to obtain high-quality wavefronts. Abaie \textit {et al}.~\cite{Mafi-Behnam-Optica-2018}
showed, in agreement with the theoretical analysis of Ref.~\cite{Mafi-Abaie-OL-2018}, that a considerable
number of the guided modes have low M$^2$ values. These high-quality modes are distributed across the transverse profile of the 
disordered fiber and can be excited without requiring sophisticated spatial light modulators at the input facet. Alternatively,
when the input light is coupled to the entire transverse area of a TAL fiber, the output is spatially incoherent. Therefore, by
proper coupling of the light, without sophisticated spatial modulators, it is possible to access a range of spatial coherence properties
in these fibers. Of particular importance is the possibility of using part of the transverse structure of the fiber to guide
spatially incoherent light to illuminate a scene and the other parts of the fiber to transport the images back in an endoscopic setting. 
The possibility of generating incoherent but directional broadband light was also highlighted in Ref.~\cite{Mafi-Behnam-Random-Laser-2017}
in a TAL random laser setup. 
%%%%%%%%%%%%%%%%%%%%%%%%%%%%%%%%%%%%%%%%%%%%%%%%%%%%%%%%%%%%%%%%%%%%%%%%%%%%%%%%
%%%%%%%%%%%%%%%%%%%%%%%%%%%%%%%%%%%%%%%%%%%%%%%%%%%%%%%%%%%%%%%%%%%%%%%%%%%%%%%%
\section{Future directions and conclusions}
%%%%%%%%%%%%%%%%%%%%%%%%%%%%%%%%%%%%%%%%%%%%%%%%%%%%%%%%%%%%%%%%%%%%%%%%%%%%%%%%
%%%%%%%%%%%%%%%%%%%%%%%%%%%%%%%%%%%%%%%%%%%%%%%%%%%%%%%%%%%%%%%%%%%%%%%%%%%%%%%%
Disordered optical fibers demonstrate many novel physical properties, mainly driven by the possibility of the 
localized beam transport over the entire cross section of the optical fiber. Therefore, they should be designed
to support highly localized modes with small localization radii, in order to have a narrow 
point-spread function for high-resolution image transport. Localized modes must also be uniformly distributed 
over the transverse cross section of the fiber, and the mode-to-mode variations in the localization radius 
must be minimized to improve the image transport uniformity. At present, these requirements translate into 
efforts in fabricating fibers with highly disordered transverse refractive index profiles with small transverse 
index correlations. In order to improve the distances over which high-quality images can be transported, it is 
imperative to ensure a high degree of longitudinal uniformity in these fibers. We anticipate that over the 
next few years, an expanded selection of materials and improved fabrication methods would enhance and expand 
the reach of these fibers, both in physical properties and in practical applications.
\section*{Acknowledgment}
A. Mafi acknowledges support by Grant Number 1807857 from National Science Foundation (NSF).
J. Ballato acknowledges support by Grant Number 1808232 from NSF.


%\ifCLASSOPTIONcaptionsoff
 % \newpage
%\fi

\providecommand{\noopsort}[1]{}\providecommand{\singleletter}[1]{#1}%




%\bibliography{anderson-refs}

\begin{thebibliography}{10}
\providecommand{\url}[1]{#1}
\csname url@samestyle\endcsname
\providecommand{\newblock}{\relax}
\providecommand{\bibinfo}[2]{#2}
\providecommand{\BIBentrySTDinterwordspacing}{\spaceskip=0pt\relax}
\providecommand{\BIBentryALTinterwordstretchfactor}{4}
\providecommand{\BIBentryALTinterwordspacing}{\spaceskip=\fontdimen2\font plus
\BIBentryALTinterwordstretchfactor\fontdimen3\font minus
  \fontdimen4\font\relax}
\providecommand{\BIBforeignlanguage}[2]{{%
\expandafter\ifx\csname l@#1\endcsname\relax
\typeout{** WARNING: IEEEtran.bst: No hyphenation pattern has been}%
\typeout{** loaded for the language `#1'. Using the pattern for}%
\typeout{** the default language instead.}%
\else
\language=\csname l@#1\endcsname
\fi
#2}}
\providecommand{\BIBdecl}{\relax}
\BIBdecl

\bibitem{Anderson1}
P.~W. Anderson, ``Absence of diffusion in certain random lattices,''
  \emph{Phys. Rev.}, vol. 109, no.~5, pp. 1492--1505, Mar. 1958.

\bibitem{Anderson1980}
P.~W. Anderson, D.~J. Thouless, E.~Abrahams, and D.~S. Fisher, ``New method for
  a scaling theory of localization,'' \emph{Phys. Rev. B}, vol.~22, no.~8, pp.
  3519--3526, Oct. 1980.

\bibitem{Abrahams-50-book}
E.~Abrahams, \emph{50 years of {A}nderson Localization}.\hskip 1em plus 0.5em
  minus 0.4em\relax Singapore: world scientific, 2010.

\bibitem{Lagendijk-Physics-Today-2009}
A.~Lagendijk, B.~van Tiggelen, and D.~S. Wiersma, ``Fifty-years of {A}nderson
  localization,'' \emph{Physics Today}, vol.~62, no.~8, pp. 24--29, Aug. 2009.

\bibitem{Abrahams-Scaling-Theory}
E.~Abrahams, P.~W. Anderson, D.~C. Licciardello, and T.~V. Ramakrishnan,
  ``Scaling theory of localization: Absence of quantum diffusion in two
  dimensions,'' \emph{Phys. Rev. Lett.}, vol.~42, no.~10, pp. 673--676, Mar.
  1979.

\bibitem{Stone}
A.~D. Stone and J.~D. Joannopoulos, ``Probability distribution and new scaling
  law for the resistance of a one-dimensional {A}nderson model,'' \emph{Phys.
  Rev. B}, vol.~24, no.~6, pp. 3592--3595, Sep 1981.

\bibitem{sheng2006introduction}
P.~Sheng, \emph{Introduction to wave scattering, localization and mesoscopic
  phenomena}, 2nd~ed.\hskip 1em plus 0.5em minus 0.4em\relax Berlin, Germany:
  Springer-Verlag, 2006.

\bibitem{Thouless-1974}
D.~J. Thouless, ``Electrons in disordered systems and the theory of
  localization,'' \emph{Physics Reports}, vol.~13, no.~3, pp. 93--142, Oct.
  1974.

\bibitem{Wegner1976}
F.~J. Wegner, ``Electrons in disordered systems. scaling near the mobility
  edge,'' \emph{Zeitschrift f{\"u}r Physik B Condensed Matter}, vol.~25, no.~4,
  pp. 327--337, Dec. 1976.

\bibitem{Soukoulis-1999}
C.~M. Soukoulis and E.~N. Economou, ``Electronic localization in disordered
  systems,'' \emph{Waves in Random Media}, vol.~9, no.~2, pp. 255--269, Dec.
  1999.

\bibitem{matter-waves-2008}
J.~Billy, V.~Josse, Z.~Zuo, A.~Bernard, B.~Hambrecht, P.~Lugan, D.~Cl\'{e}ment,
  L.~Sanchez-Palencia, P.~Bouyer, and A.~Aspect, ``Direct observation of
  {A}nderson localization of matter waves in a controlled disorder,''
  \emph{Nature}, vol. 453, no. 7197, pp. 891--894, Jun. 2008.

\bibitem{roati2008anderson}
G.~Roati, C.~D'Errico, L.~Fallani, M.~Fattori, C.~Fort, M.~Zaccanti,
  G.~Modugno, M.~Modugno, and M.~Inguscio, ``{A}nderson localization of a
  non-interacting {B}ose--{E}instein condensate,'' \emph{Nature}, vol. 453, no.
  7197, p. 895, Jun. 2008.

\bibitem{kondov2011three}
S.~Kondov, W.~McGehee, J.~Zirbel, and B.~DeMarco, ``Three-dimensional
  {A}nderson localization of ultracold matter,'' \emph{Science}, vol. 334, no.
  6052, pp. 66--68, Oct. 2011.

\bibitem{quantum-fields-2010}
C.~Thompson, G.~Vemuri, and G.~S. Agarwal, ``{A}nderson localization with
  second quantized fields in a coupled array of waveguides,'' \emph{Phys. Rev.
  A}, vol.~82, no.~5, p. 053805, Nov. 2010.

\bibitem{Lahini-Quantum-Correlation-2010}
Y.~Lahini, Y.~Bromberg, D.~N. Christodoulides, and Y.~Silberberg, ``Quantum
  correlations in two-particle {A}nderson localization,'' \emph{Phys. Rev.
  Lett.}, vol. 105, no.~16, p. 163905, Oct. 2010.

\bibitem{Lahini-HBT-2011}
Y.~Lahini, Y.~Bromberg, Y.~Shechtman, A.~Szameit, D.~N. Christodoulides,
  R.~Morandotti, and Y.~Silberberg, ``Hanbury {B}rown and {T}wiss correlations
  of {A}nderson localized waves,'' \emph{Phys. Rev. A}, vol.~84, no.~4, p.
  041806, Oct. 2011.

\bibitem{Abouraddy-entangled-2012}
A.~F. Abouraddy, G.~Di~Giuseppe, D.~N. Christodoulides, and B.~E.~A. Saleh,
  ``{A}nderson localization and colocalization of spatially entangled
  photons,'' \emph{Phys. Rev. A}, vol.~86, no.~4, p. 040302, Oct. 2012.

\bibitem{ultrasound-1990}
R.~Weaver, ``{A}nderson localization of ultrasound,'' \emph{Wave Motion},
  vol.~12, no.~2, pp. 129--142, Mar. 1990.

\bibitem{acoustic-PRL-1990}
I.~S. Graham, L.~Pich\'e, and M.~Grant, ``Experimental evidence for
  localization of acoustic waves in three dimensions,'' \emph{Phys. Rev.
  Lett.}, vol.~64, no.~26, pp. 3135--3138, Jun. 1990.

\bibitem{elastics-Nat-Phys-2009}
H.~Hu, A.~Strybulevych, J.~H. Page, S.~E. Skipetrov, and B.~A. van Tiggelen,
  ``Localization of ultrasound in a three-dimensional elastic network,''
  \emph{Nat. Phys.}, vol.~4, no.~12, pp. 945--948, Dec. 2008.

\bibitem{John-EM-abs-mobility-edge-1984}
S.~John, ``Electromagnetic absorption in a disordered medium near a photon
  mobility edge,'' \emph{Phys. Rev. Lett.}, vol.~53, no.~22, pp. 2169--2172,
  Nov. 1984.

\bibitem{dalichaouch1991microwave}
R.~Dalichaouch, J.~Armstrong, S.~Schultz, P.~Platzman, and S.~McCall,
  ``Microwave localization by two-dimensional random scattering,''
  \emph{Nature}, vol. 354, no. 6348, p.~53, 1991.

\bibitem{Chabanov-microwave-2000}
{A.~A.~Chabanov, M.~Stoytchev, A.~Z.~Genack}, ``Statistical signatures of
  photon localization,'' \emph{Nature}, vol. 404, no. 6780, pp. 850--853, Apr.
  2000.

\bibitem{El-Dardiry-microwave-2012}
R.~G.~S. El-Dardiry, S.~Faez, and A.~Lagendijk, ``Snapshots of {A}nderson
  localization beyond the ensemble average,'' \emph{Phys. Rev. B}, vol.~86,
  no.~12, p. 125132, Sep. 2012.

\bibitem{Anderson2}
P.~W. Anderson, ``The question of classical localization a theory of white
  paint?'' \emph{Philos. Mag. B}, vol.~52, no.~3, pp. 505--509, Sep. 1985.

\bibitem{John-photon-localization-1987}
S.~John, ``Strong localization of photons in certain disordered dielectric
  superlattices,'' \emph{Phys. Rev. Lett.}, vol.~58, no.~23, pp. 2486--2489,
  Jun. 1987.

\bibitem{John-Physics-Today-1991}
{S.~John}, ``Localization of light,'' \emph{Phys. Today}, vol.~44, no.~5, pp.
  32--40, 1991.

\bibitem{SegevNaturePhotonicsReview}
Y.~S. Mordechai~Segev and D.~N. Christodoulides, ``Anderson localization of
  light,'' \emph{Nat. Photonics}, vol.~7, no.~3, pp. 197--204, Feb. 2013.

\bibitem{Mafi-AOP-2015}
A.~Mafi, ``Transverse {A}nderson localization of light: a tutorial,''
  \emph{Adv. Opt. Photon.}, vol.~7, no.~3, pp. 459--515, Sep. 2015.

\bibitem{storzer2006observation}
M.~St{\"o}rzer, P.~Gross, C.~M. Aegerter, and G.~Maret, ``Observation of the
  critical regime near {A}nderson localization of light,'' \emph{Phys. Rev.
  Lett.}, vol.~96, no.~6, p. 063904, 2006.

\bibitem{wiersma1997localization}
D.~S. Wiersma, P.~Bartolini, A.~Lagendijk, and R.~Righini, ``Localization of
  light in a disordered medium,'' \emph{Nature}, vol. 390, no. 6661, p. 671,
  1997.

\bibitem{yannopapas2003anderson}
V.~Yannopapas, A.~Modinos, and N.~Stefanou, ``{A}nderson localization of light
  in inverted opals,'' \emph{Phys. Rev. B}, vol.~68, no.~19, p. 193205, 2003.

\bibitem{aegerter2007observation}
C.~M. Aegerter, M.~St{\"o}rzer, S.~Fiebig, W.~B{\"u}hrer, and G.~Maret,
  ``Observation of {A}nderson localization of light in three dimensions,''
  \emph{J. Opt. Soc. Am. A}, vol.~24, no.~10, pp. A23--A27, 2007.

\bibitem{Lahini-1D-AL-2008}
Y.~Lahini, A.~Avidan, F.~Pozzi, M.~Sorel, R.~Morandotti, D.~N. Christodoulides,
  and Y.~Silberberg, ``{A}nderson localization and nonlinearity in
  one-dimensional disordered photonic lattices,'' \emph{Phys. Rev. Lett.}, vol.
  100, no.~1, p. 013906, Jan. 2008.

\bibitem{fishman2012nonlinear}
S.~Fishman, Y.~Krivolapov, and A.~Soffer, ``The nonlinear schr{\"o}dinger
  equation with a random potential: results and puzzles,'' \emph{Nonlinearity},
  vol.~25, no.~4, p. R53, 2012.

\bibitem{mafi-NL-ArXiv-2017}
A.~Mafi, ``A brief overview of the interplay between nonlinearity and
  transverse {A}nderson localization,'' \emph{arXiv:1703.04011}, 2017.

\bibitem{Mafi-Marco-PRL-Migrating-NL-2014}
M.~Leonetti, S.~Karbasi, A.~Mafi, and C.~Conti, ``Observation of migrating
  transverse {A}nderson localizations of light in nonlocal media,'' \emph{Phys.
  Rev. Lett.}, vol. 112, no.~19, p. 193902, May 2014.

\bibitem{Mafi-Marco-APL-self-focusing-2014}
{Leonetti, Marco and Karbasi, Salman and Mafi, Arash and Conti, Claudio},
  ``Experimental observation of disorder induced self-focusing in optical
  fibers,'' \emph{Appl. Phys. Lett.}, vol. 105, no.~17, p. 171102, Oct. 2014.

\bibitem{sperling2013direct}
T.~Sperling, W.~Buehrer, C.~M. Aegerter, and G.~Maret, ``Direct determination
  of the transition to localization of light in three dimensions,'' \emph{Nat.
  Photonics}, vol.~7, no.~1, p.~48, 2013.

\bibitem{vatnik2017anderson}
I.~D. Vatnik, A.~Tikan, G.~Onishchukov, D.~V. Churkin, and A.~A. Sukhorukov,
  ``{A}nderson localization in synthetic photonic lattices,'' \emph{Sci. Rep.},
  vol.~7, no.~1, p. 4301, 2017.

\bibitem{choi2018anderson}
S.~H. Choi, S.-W. Kim, Z.~Ku, M.~A. Visbal-Onufrak, S.-R. Kim, K.-H. Choi,
  H.~Ko, W.~Choi, A.~M. Urbas, T.-W. Goo \emph{et~al.}, ``{A}nderson light
  localization in biological nanostructures of native silk,'' \emph{Nat.
  Commun.}, vol.~9, no.~1, p. 452, 2018.

\bibitem{transverse-Abdullaev}
S.~S. Abdullaev and F.~K. Abdullaev, ``On propagation of light in fiber bundles
  with random parameters,'' \emph{Radiofizika}, vol.~23, no.~6, pp. 766--767,
  1980.

\bibitem{transverse-DeRaedt}
H.~De~Raedt, A.~Lagendijk, and P.~de~Vries, ``Transverse localization of
  light,'' \emph{Phys. Rev. Lett.}, vol.~62, no.~1, pp. 47--50, Jan. 1989.

\bibitem{Saleh-Teich}
B.~E.~A. Saleh and M.~C. Teich, \emph{{Fundamentals of photonics; 2nd ed.}},
  ser. Wiley series in pure and applied optics.\hskip 1em plus 0.5em minus
  0.4em\relax New York, NY: Wiley, 2007.

\bibitem{Pertsch}
T.~Pertsch, U.~Peschel, J.~Kobelke, K.~Schuster, H.~Bartelt, S.~Nolte,
  A.~T\"unnermann, and F.~Lederer, ``Nonlinearity and disorder in fiber
  arrays,'' \emph{Phys. Rev. Lett.}, vol.~93, no.~5, p. 053901, Jul. 2004.

\bibitem{Schwartz2007}
T.~Schwartz, G.~Bartal, S.~Fishman, and M.~Segev, ``Transport and {A}nderson
  localization in disordered two-dimensional photonic lattices,''
  \emph{Nature}, vol. 446, no. 7131, pp. 52--55, Mar. 2007.

\bibitem{Szameit-boundary-2010}
A.~Szameit, Y.~V. Kartashov, P.~Zeil, F.~Dreisow, M.~Heinrich, R.~Keil,
  S.~Nolte, A.~T\"{u}nnermann, V.~Vysloukh, and L.~Torner, ``Wave localization
  at the boundary of disordered photonic lattices,'' \emph{Opt. Lett.},
  vol.~35, no.~8, pp. 1172--1174, Apr. 2010.

\bibitem{Martin-1D-AL-2011}
L.~Martin, G.~D. Giuseppe, A.~Perez-Leija, R.~Keil, F.~Dreisow, M.~Heinrich,
  S.~Nolte, A.~Szameit, A.~F. Abouraddy, D.~N. Christodoulides, and B.~E.~A.
  Saleh, ``{A}nderson localization in optical waveguide arrays with
  off-diagonal coupling disorder,'' \emph{Opt. Express}, vol.~19, no.~14, pp.
  13\,636--13\,646, Jul. 2011.

\bibitem{Kartashov-NL-2012}
Y.~V. Kartashov, V.~V. Konotop, V.~A. Vysloukh, and L.~Torner, ``Light
  localization in nonuniformly randomized lattices,'' \emph{Opt. Lett.},
  vol.~37, no.~3, pp. 286--288, Feb. 2012.

\bibitem{Mafi-Salman-OL-2012}
S.~Karbasi, C.~R. Mirr, P.~G. Yarandi, R.~J. Frazier, K.~W. Koch, and A.~Mafi,
  ``Observation of transverse {A}nderson localization in an optical fiber,''
  \emph{Opt. Lett.}, vol.~37, no.~12, pp. 2304--2306, Jun. 2012.

\bibitem{Mafi-Salman-JOVE-2013}
S.~Karbasi, R.~J. Frazier, C.~R. Mirr, K.~W. Koch, and A.~Mafi, ``Fabrication
  and characterization of disordered polymer optical fibers for transverse
  {A}nderson localization of light,'' \emph{J. Vis. Exp.}, vol.~77, p. e50679,
  Jul. 2013.

\bibitem{Mafi-Salman-OPEX-2012}
S.~Karbasi, C.~R. Mirr, R.~J. Frazier, P.~G. Yarandi, K.~W. Koch, and A.~Mafi,
  ``Detailed investigation of the impact of the fiber design parameters on the
  transverse {A}nderson localization of light in disordered optical fibers,''
  \emph{Opt. Express}, vol.~20, no.~17, pp. 18\,692--18\,706, Aug. 2012.

\bibitem{Skipetrov}
S.~E. Skipetrov and I.~M. Sokolov, ``Absence of {A}nderson localization of
  light in a random ensemble of point scatterers,'' \emph{Phys. Rev. Lett.},
  vol. 112, no.~2, p. 023905, Jan. 2014.

\bibitem{Mafi-Salman-OMEX-2012}
S.~Karbasi, T.~Hawkins, J.~Ballato, K.~W. Koch, and A.~Mafi, ``Transverse
  {A}nderson localization in a disordered glass optical fiber,'' \emph{Opt.
  Mater. Express}, vol.~2, no.~11, pp. 1496--1503, Nov. 2012.

\bibitem{Jovic-boundary-2011}
D.~M. Jovi\'{c}, Y.~S. Kivshar, C.~Denz, and M.~R. Beli\'{c}, ``{A}nderson
  localization of light near boundaries of disordered photonic lattices,''
  \emph{Phys. Rev. A}, vol.~83, no.~3, p. 033813, Mar. 2011.

\bibitem{Molina-boundary-2011}
M.~I. Molina, ``Boundary-induced {A}nderson localization in photonic
  lattices,'' \emph{Phys. Lett. A}, vol. 375, no.~20, pp. 2056--2058, 2011.

\bibitem{Naether-boundary-2012}
U.~Naether, J.~M. Meyer, S.~St\"{u}tzer, A.~T\"{u}nnermann, S.~Nolte, M.~I.
  Molina, and A.~Szameit, ``{A}nderson localization in a periodic photonic
  lattice with a disordered boundary,'' \emph{Opt. Lett.}, vol.~37, no.~4, pp.
  485--487, Feb. 2012.

\bibitem{Mafi-Behnam-Boundary-OC-2016}
B.~Abaie, S.~R. Hosseini, S.~Karbasi, and A.~Mafi, ``Modal analysis of the
  impact of the boundaries on transverse {A}nderson localization in a
  one-dimensional disordered optical lattice,'' \emph{Opt. Comm.}, vol. 365,
  pp. 208--214, Apr. 2016.

\bibitem{chen2014observing}
M.~Chen and M.-J. Li, ``Observing transverse {A}nderson localization in random
  air line based fiber,'' in \emph{Photonic and Phononic Properties of
  Engineered Nanostructures IV}, vol. 8994.\hskip 1em plus 0.5em minus
  0.4em\relax International Society for Optics and Photonics, 2014, p. 89941S.

\bibitem{ZHAO:17}
J.~Zhao, J.~E. Antonio-Lopez, R.~A. Correa, A.~Mafi, M.~Windeck, and
  A.~Sch\"{u}lzgen, ``Image transport through silica-air random core optical
  fiber,'' in \emph{Conference on Lasers and Electro-Optics}.\hskip 1em plus
  0.5em minus 0.4em\relax Optical Society of America, 2017, p. JTu5A.91.

\bibitem{zhao2018image}
J.~Zhao, J.~E.~A. Lopez, Z.~Zhu, D.~Zheng, S.~Pang, R.~A. Correa, and
  A.~Sch{\"u}lzgen, ``Image transport through meter-long randomly disordered
  silica-air optical fiber,'' \emph{Sci. Rep.}, vol.~8, no.~1, p. 3065, 2018.

\bibitem{tong2018characterization}
H.~T. Tong, S.~Kuroyanagi, K.~Nagasaka, T.~Suzuki, and Y.~Ohishi,
  ``Characterization of an all-solid disordered tellurite glass optical fiber
  and its near-infrared optical image transport,'' \emph{Jpn. J. Appl. Phys.},
  Dec. 2018.

\bibitem{Mafi-Salman-Modal-JOSAB-2013}
S.~Karbasi, K.~W. Koch, and A.~Mafi, ``Modal perspective on the transverse
  {A}nderson localization of light in disordered optical lattices,'' \emph{J.
  Opt. Soc. Am. B}, vol.~30, no.~6, pp. 1452--1461, Jun. 2013.

\bibitem{Mafi-Schirmacher-PRL-2018}
W.~Schirmacher, B.~Abaie, A.~Mafi, G.~Ruocco, and M.~Leonetti, ``What is the
  right theory for {A}nderson localization of light? an experimental test,''
  \emph{Phys. Rev. Lett.}, vol. 120, no.~6, p. 067401, Feb. 2018.

\bibitem{Mafi-Salman-Multiple-Beam-2013}
S.~Karbasi, K.~W. Koch, and A.~Mafi, ``Multiple-beam propagation in an
  {A}nderson localized optical fiber,'' \emph{Opt. Express}, vol.~21, no.~1,
  pp. 305--313, Jan. 2013.

\bibitem{Mafi-Salman-Image-2013}
{S. Karbasi, K. W. Koch, and A. Mafi}, ``Image transport quality can be
  improved in disordered waveguides,'' \emph{Opt. Comm.}, vol. 311, pp. 72--76,
  Jan. 2013.

\bibitem{Mafi-Salman-Nature-2014}
S.~Karbasi, R.~J. Frazier, K.~W. Koch, T.~Hawkins, J.~Ballato, and A.~Mafi,
  ``Image transport through a disordered optical fibre mediated by transverse
  {A}nderson localization,'' \emph{Nat Commun}, vol.~5, no. 3362, Feb. 2014.

\bibitem{Mafi-Marco-information-2016}
M.~Leonetti, S.~Karbasi, A.~Mafi, E.~DelRe, and C.~Conti, ``Secure information
  transport by transverse localization of light,'' \emph{Sci. Rep.}, vol.~6,
  no. 29918, Jul. 2016.

\bibitem{zhao2018deep}
J.~Zhao, Y.~Sun, Z.~Zhu, J.~E. Antonio-Lopez, R.~A. Correa, S.~Pang, and
  A.~Schülzgen, ``Deep learning imaging through fully-flexible glass-air
  disordered fiber,'' \emph{ACS Photonics}, vol.~5, no.~10, pp. 3930--3935,
  2018.

\bibitem{Choi-multimode}
Y.~Choi, C.~Yoon, M.~Kim, T.~D. Yang, C.~Fang-Yen, R.~R. Dasari, K.~J. Lee, and
  W.~Choi, ``Scanner-free and wide-field endoscopic imaging by using a single
  multimode optical fiber,'' \emph{Phys. Rev. Lett.}, vol. 109, no.~20, p.
  203901, Nov. 2012.

\bibitem{seward1974elongation}
T.~P. Seward~III, ``Elongation and spheroidization of phase-separated particles
  in glass,'' \emph{J. Non-Cryst. Solids}, vol.~15, no.~3, pp. 487--504, 1974.

\bibitem{seward1977some}
T.~Seward~III, ``Some unusual optical properties of elongated phases in
  glasses,'' \emph{The Physics of Non-Crystalline Solids; Trans Tech
  Publications: Aedermannsdorf, Switzerland}, pp. 342--347, 1977.

\bibitem{Pichard1981-1}
J.~L. Pichard and G.~Sarma, ``Finite size scaling approach to {A}nderson
  localisation,'' \emph{J. Phys. C. Solid State Phys.}, vol.~14, no.~6, p.
  L127, 1981.

\bibitem{Pichard1981-2}
{J. L. Pichard and G. Sarma}, ``Finite-size scaling approach to {A}nderson
  localisation. ii. quantitative analysis and new results,'' \emph{J. Phys. C.
  Solid State Phys.}, vol.~14, no.~21, p. L617, 1981.

\bibitem{Pichard1986-1}
J.~L. Pichard, ``The one-dimensional {A}nderson model: scaling and resonances
  revisited,'' \emph{J. Phys. C. Solid State Phys.}, vol.~19, no.~10, p. 1519,
  1986.

\bibitem{Pichard1986-2}
J.~L. Pichard and G.~Andr\'{e}, ``Many-channel transmission: Large volume limit
  of the distribution of localization lengths and one-parameter scaling,''
  \emph{EPL}, vol.~2, no.~6, p. 477, 1986.

\bibitem{Aegerter:07}
C.~M. Aegerter, M.~St\"{o}rzer, S.~Fiebig, W.~B\"{u}hrer, and G.~Maret,
  ``Scaling behavior of the {A}nderson localization transition of light,'' in
  \emph{Photonic Metamaterials: From Random to Periodic}.\hskip 1em plus 0.5em
  minus 0.4em\relax Optical Society of America, 2007, p. TuA2.

\bibitem{Mafi-Behnam-Scaling-PRB-2016}
B.~Abaie and A.~Mafi, ``Scaling analysis of transverse {A}nderson localization
  in a disordered optical waveguide,'' \emph{Phys. Rev. B}, vol.~94, no.~6, p.
  064201, Aug. 2016.

\bibitem{Mafi-Abaie-OL-2018}
{B. Abaie and A. Mafi}, ``Modal area statistics for transverse {A}nderson
  localization in disordered optical fibers,'' \emph{Opt. Lett.}, vol.~43,
  no.~16, pp. 3834--3837, Aug. 2018.

\bibitem{Mafi-Marco-singlemode-2017}
G.~Ruocco, B.~Abaie, W.~Schirmacher, A.~Mafi, and M.~Leonetti,
  ``Disorder-induced single-mode transmission,'' \emph{Nat. Commun.}, vol.~8,
  no. 14571, Mar. 2017.

\bibitem{Mafi-Behnam-Optica-2018}
B.~Abaie, M.~Peysokhan, J.~Zhao, J.~E. Antonio-Lopez, R.~Amezcua-Correa,
  A.~Sch\"{u}lzgen, and A.~Mafi, ``Disorder-induced high-quality wavefront in
  an {A}nderson localizing optical fiber,'' \emph{Optica}, vol.~5, no.~8, pp.
  984--987, Aug. 2018.

\bibitem{Mafi-Behnam-Random-Laser-2017}
B.~Abaie, E.~Mobini, S.~Karbasi, T.~Hawkins, J.~Ballato, and A.~Mafi, ``Random
  lasing in an {A}nderson localizing optical fiber,'' \emph{Light Sci. Appl.},
  vol.~6, no. e17041, Aug. 2017.

\end{thebibliography}
%\bibliographystyle{IEEEtran}


\end{document}